\documentclass[11pt,letterpaper]{article}
\usepackage[utf8]{inputenc}
\usepackage[margin=1in]{geometry}
\usepackage{times}
\usepackage{amsmath,amssymb,amsfonts}
\usepackage{graphicx}
\usepackage{booktabs}
\usepackage{multirow}
\usepackage{array}
\usepackage{url}
\usepackage{hyperref}
\usepackage{setspace}
\usepackage{titlesec}
\usepackage{fancyhdr}
\usepackage{enumitem}

\titleformat{\section}{\large\bfseries}{\thesection.}{0.5em}{}
\titleformat{\subsection}{\normalfont\bfseries}{\thesubsection.}{0.5em}{}
\titleformat{\subsubsection}{\normalfont\itshape}{\thesubsubsection.}{0.5em}{}

\begin{document}

% =========================================================================
% TITLE
% =========================================================================
\begin{center}
    {\LARGE \bf AVNet} \\[0.5em]
    {\Large Multimodal Emergency Vehicle Classification via\\
    Audio-Visual Transformers and Knowledge Distillation} \\[1em]
    \textbf{Vijay John \quad Amar Dabaja} \\[0.3em]
    Department of Mathematics and Computer Science \\
    Lawrence Technological University, Southfield, MI, USA \\[0.3em]
    \texttt{\{vjohn, adabaja\}@ltu.edu} \\[1em]
    \textit{March 2026}
\end{center}

\vspace{1em}
\begin{abstract}
Emergency vehicle detection in autonomous driving is a safety-critical perception
task that demands robustness under diverse and adverse real-world conditions. Existing
approaches rely on a single modality, either audio or video, which leads to
systematic failure when that modality is degraded: microphone-based systems fail in
noisy urban environments, and camera-based systems fail at night or under occlusion.
This report presents \textbf{AVNet}, a multimodal audio-visual transformer that
classifies emergency vehicles (ambulance, fire engine, police car) and road background
using both audio and video, while gracefully handling the absence of either modality
at inference time. AVNet introduces three key contributions: (1) a \textit{temporally
aligned cross-modal fusion} module that performs second-level cross-attention between
audio spectrogram tokens and video frame tokens, exploiting their exact temporal
correspondence without any learned alignment mechanism; (2) \textit{learned null
embeddings} that substitute for missing modality tokens, enabling a single unified
model to operate in audio-only, video-only, or joint audio-visual mode without
retraining; and (3) a \textit{knowledge distillation} training strategy in which
specialist unimodal teacher models transfer inter-class dark knowledge into the
multimodal student fusion branch via soft probability targets. Evaluated on 281 clips
from the Google AudioSet dataset, AVNet achieves 66.6\% overall accuracy in
audio-visual mode, outperforming the audio-only branch by $+10.4\%$ and the
video-only branch by $+15.0\%$. The largest per-class gain is observed for the
hardest class, Ambulance, where fusion achieves $+29.5\%$ over either unimodal
branch alone, demonstrating that the two modalities provide complementary information
that the aligned cross-attention mechanism successfully exploits.
\end{abstract}

% =========================================================================
% SECTION 1: INTRODUCTION
% =========================================================================
\section{Introduction}

Emergency vehicle detection is a critical task in intelligent transportation systems
and autonomous driving. When an ambulance, fire engine, or police car approaches,
nearby vehicles must yield immediately. While humans rely on both the sound of a siren
and the visual appearance of flashing lights, automated systems often use only one
modality, either audio or video, which leads to higher error rates under adverse
conditions. A microphone-based system fails in noisy urban environments; a
camera-based system fails at night or when the vehicle is occluded. Neither modality
alone is sufficient for robust real-world detection.

This report presents \textbf{AVNet}, a multimodal deep learning model that classifies
emergency vehicles using audio and/or video. The system is designed to handle three
real-world scenarios:
\begin{enumerate}
    \item \textbf{Audio-only:} microphone available, no camera (or obstructed view)
    \item \textbf{Video-only:} camera available, no microphone (or high noise
    environment)
    \item \textbf{Audio-Video:} both modalities available
\end{enumerate}

The model is trained on the AudioSet dataset and uses a student-teacher training
strategy called Knowledge Distillation to improve the fusion branch by learning from
specialist unimodal models. The key design contributions of AVNet are: (1) a
temporally aligned cross-modal fusion module that exploits the second-level
correspondence between audio spectrogram tokens and video frame tokens; (2) learned
null embeddings that allow the model to gracefully handle missing modalities at
inference time without retraining; and (3) a knowledge distillation training strategy
that transfers dark knowledge from specialist teacher models into the multimodal
student.

The remainder of this report is structured as follows. Section~2 describes the
proposed AVNet framework in full, covering the model architecture, knowledge
distillation strategy, and training pipeline. Section~3 describes the AudioSet
dataset and the audio and video preprocessing pipelines. Section~4 presents
experimental results and analysis. Section~5 discusses strengths, limitations, and
future directions. Section~6 concludes.

% =========================================================================
% SECTION 2: PROPOSED FRAMEWORK
% =========================================================================
\section{Proposed Framework}

\subsection{Model Architecture}

\begin{figure}[t]
\centering
\includegraphics[width=0.75\linewidth]{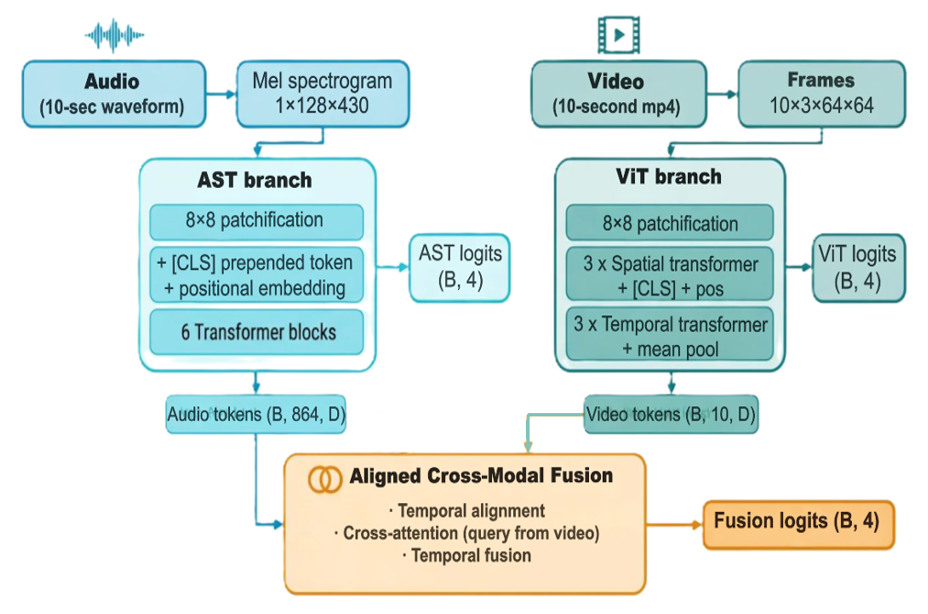}
\caption{AVNet architecture. The AST branch processes audio spectrograms and the ViT
branch processes video frames. Both produce token sequences that are fused via
temporally aligned cross-attention. Dashed arrows indicate branch-specific logits used
during training. Missing modalities are replaced by learned null embeddings inside the
fusion module, enabling the model to produce valid outputs regardless of which
modalities are present at inference time.}
\label{fig:avnet_arch}
\end{figure}

AVNet consists of three components: the \textbf{AST Branch} (audio encoder), the
\textbf{ViT Branch} (video encoder), and the \textbf{Aligned Cross-Modal Fusion}
module. Each branch independently processes its modality and produces both a set of
logits for unimodal classification and a sequence of tokens that are passed to the
fusion module. The full model is shown schematically in Figure~\ref{fig:avnet_arch}.

\subsubsection{The Transformer Block}

All three components of AVNet are built from the same fundamental unit: the
Transformer block~\cite{vaswani2017attention}. Understanding this block is essential
for understanding the entire architecture. Given an input sequence $\mathbf{X} \in
\mathbb{R}^{N \times D}$ of $N$ tokens each with dimension $D$, the Transformer
block applies two sub-layers with residual connections and layer normalisation.

\textbf{Multi-Head Self-Attention (MHSA)} allows each token to attend to all other
tokens in the sequence, capturing global dependencies:
\begin{equation}
    \text{MHSA}(\mathbf{X}) = \text{Concat}(\text{head}_1, \ldots,
    \text{head}_h)\mathbf{W}^O
\end{equation}
where each attention head computes a scaled dot-product attention:
\begin{equation}
    \text{head}_i = \text{Attention}(\mathbf{X}\mathbf{W}^Q_i, \mathbf{X}\mathbf{W}^K_i,
    \mathbf{X}\mathbf{W}^V_i), \qquad
    \text{Attention}(Q,K,V) = \text{softmax}\!\left(\frac{QK^\top}{\sqrt{d_k}}\right)V
\end{equation}
The division by $\sqrt{d_k}$ prevents the dot products from growing too large as the
key dimension $d_k$ increases, which would push the softmax into regions of very
small gradients and destabilise training.

\textbf{Feed-Forward Network (MLP)} applies a position-wise two-layer network with
GELU activation:
\begin{equation}
    \text{MLP}(x) = \mathbf{W}_2 \cdot \text{GELU}(\mathbf{W}_1 x + b_1) + b_2
\end{equation}
with hidden dimension $= 4D$ (default in all AVNet components).

The full block uses pre-layer normalisation and residual connections, which have been
shown to improve training stability over the original post-norm design:
\begin{align}
    \mathbf{X}' &= \mathbf{X} + \text{MHSA}(\text{LayerNorm}(\mathbf{X})) \\
    \mathbf{X}'' &= \mathbf{X}' + \text{MLP}(\text{LayerNorm}(\mathbf{X}'))
\end{align}
In the fusion module, cross-attention is used instead of self-attention: the query
comes from the video tokens and the keys and values come from the audio tokens, so
each video token attends selectively to the audio tokens from the corresponding
time window.

\subsubsection{AST Branch: Audio Spectrogram Transformer}

The Audio Spectrogram Transformer (AST)~\cite{gong2021ast} branch treats the log-mel
spectrogram as a 2D image and applies a Vision Transformer architecture to extract
audio features. This approach avoids the inductive biases of convolutional networks
and allows the model to capture long-range frequency and temporal dependencies
directly from the raw spectrogram representation.

\paragraph{Patchification:}
The mel spectrogram of shape $(1, 128, 430)$ is first padded along the time axis to
432 samples (the nearest multiple of 8) and then divided into non-overlapping $8
\times 8$ patches:
\begin{equation}
    N_f = 128/8 = 16 \text{ (frequency patches)}, \quad
    N_t = 432/8 = 54 \text{ (time patches)}, \quad
    N_{\text{audio}} = N_f \times N_t = 864 \text{ tokens}
\end{equation}
Each patch is a $1 \times 8 \times 8 = 64$-dimensional vector which is projected to
the model dimension $D$ by a learned linear layer. This is analogous to the patch
embedding used in Vision Transformers for images.

\paragraph{Token Layout and Temporal Indexing:}
Tokens are stored in row-major order (time outer, frequency inner):
\begin{equation}
    \text{token}[t \cdot N_f + f] = \text{patch at time patch } t,
    \text{ frequency patch } f
\end{equation}
A critical design decision is that the audio tokens for video second $i$ are
exactly at indices $[80i : 80(i+1)]$ (5 time patches $\times$ 16 frequency patches
$= 80$ tokens per second). This predictable layout makes exact temporal alignment
with video frames possible without any learned alignment mechanism.

\paragraph{Transformer Encoding:}
The encoding procedure follows the standard ViT pipeline:
\begin{enumerate}
    \item A learnable \texttt{[CLS]} token is added bringing the sequence length
    to 865.
    \item A learnable positional embedding of shape $(865, D)$ is added to encode the
    2D position of each patch.
    \item 6 Transformer blocks (self-attention) are applied to the full sequence.
    \item The \texttt{[CLS]} token representation is projected via a linear layer to
    produce logits $(B, 4)$ for unimodal audio classification.
    \item The remaining 864 patch tokens $(B, 864, D)$ are forwarded to the fusion
    module.
\end{enumerate}

\subsubsection{ViT Branch: Video Vision Transformer}

The ViT branch~\cite{dosovitskiy2021vit} encodes video by applying two stages of
transformer processing: a spatial stage that operates on individual frames, and a
temporal stage that models relationships between frames over time. Separating these
two stages allows the model to learn appearance features within each frame
independently from motion features across frames, which is more sample-efficient than
applying a single joint spatiotemporal transformer.

\paragraph{Spatial Encoding (per frame):}
Each of the 10 frames of shape $(3, 64, 64)$ is divided into $8 \times 8$ patches:
\begin{equation}
    N_p = (64/8)^2 = 64 \text{ patches per frame}, \quad
    \text{patch dimension} = 3 \times 8 \times 8 = 192
\end{equation}
A shared spatial transformer (3 blocks) processes all $B \times 10$ frames
simultaneously by treating them as a batch. This weight sharing across frames is
important: it ensures that the spatial encoder learns frame-level features that
generalise across time positions rather than overfitting to specific temporal
locations. The \texttt{[CLS]} token of each frame is extracted as a compact
frame-level representation, giving a tensor of shape $(B, 10, D)$.

\paragraph{Temporal Encoding:}
The sequence of 10 frame-level representations is then processed temporally:
\begin{enumerate}
    \item A learnable temporal positional embedding of shape $(10, D)$ is added to
    encode the temporal ordering of frames.
    \item 3 temporal Transformer blocks are applied over the sequence of 10 frame
    tokens, allowing each frame to attend to all other frames.
    \item Mean-pooling over the 10 time steps collapses the sequence to a single
    video-level representation.
    \item A linear layer projects to logits $(B, 4)$ for unimodal video classification.
    \item The token sequence $(B, 10, D)$ before mean-pooling is forwarded to the
    fusion module.
\end{enumerate}
Dividing the depth equally between spatial (3 blocks) and temporal (3 blocks) allows
the model to develop separate feature hierarchies for within-frame content and
cross-frame dynamics, which is more interpretable and has been shown to be effective
in video understanding.

\subsubsection{Aligned Cross-Modal Fusion}

The fusion module is the central contribution of AVNet. Rather than concatenating
global audio and video representations (which discards temporal structure) or using
a generic cross-attention over all tokens (which is computationally expensive and
ignores temporal correspondence), AVNet uses a \textit{temporally aligned
cross-attention}: each video frame token attends only to the audio tokens from the
same one-second time window.

\paragraph{Temporal Alignment:}
The alignment is exact and deterministic. Frame $t$ was sampled from second $t$ of
the video clip. The mel spectrogram has approximately 43 time steps per second (since
$22050 \times 1 / 512 \approx 43$), which corresponds to 5 time patches $\times$ 16
frequency patches $= 80$ audio tokens per second. Therefore, the audio tokens that
temporally correspond to video frame $t$ are exactly at indices $[80t : 80(t+1)]$ in
the token sequence. No learned alignment is required --- the alignment is built into
the preprocessing and token layout by design.

\paragraph{Cross-Attention per Second:}
For each second $t = 0, 1, \ldots, 9$, the fusion module performs:
\begin{itemize}
    \item Query: video token $\mathbf{v}_t \in \mathbb{R}^{1 \times D}$ for second $t$
    \item Key and Value: aligned audio tokens $\mathbf{a}_t \in \mathbb{R}^{80 \times D}$
    for second $t$
    \item Output: fused token $\mathbf{f}_t = \text{CrossAttn}(\mathbf{v}_t,
    \mathbf{a}_t) \in \mathbb{R}^{1 \times D}$
\end{itemize}
This produces a sequence of 10 fused tokens, each of which captures the joint
audio-visual information from a one-second window. The cross-attention operation
allows the video token to selectively attend to the most relevant frequency bands
in the audio signal at each moment in time.

\paragraph{Temporal Fusion:}
The 10 fused tokens $\{\mathbf{f}_0, \ldots, \mathbf{f}_9\}$ are stacked into a
sequence of shape $(B, 10, D)$ and processed by 4 temporal Transformer blocks.
Mean-pooling over the 10 time steps produces a single fused representation, which
is passed to a linear classifier to produce the final fusion logits $(B, 4)$.

\paragraph{Handling Missing Modalities:}
A critical practical requirement is that the system must function even when one
modality is unavailable, for example, when a camera is occluded or a microphone
fails. AVNet handles this through two learned null embeddings:
\begin{itemize}
    \item $\eta_a \in \mathbb{R}^{80 \times D}$: replaces all audio token sequences
    when audio is missing
    \item $\eta_v \in \mathbb{R}^{1 \times D}$: replaces all video tokens when video
    is missing
\end{itemize}
These null embeddings are learned parameters trained end-to-end alongside the rest
of the network. During training, modality dropout is applied stochastically: with
some probability, the audio or video tokens are replaced by their corresponding null
embeddings. This forces the model to learn meaningful null representations that allow
the fusion module to produce useful outputs even when one modality provides no
information. At inference time, the appropriate null embedding is substituted for any
missing modality, ensuring the model always produces a valid output.

\subsection{Knowledge Distillation}

\subsubsection{Motivation}

Suppose we want a single model that can handle audio-only, video-only, and
audio-video inputs simultaneously. A naive approach is to train it end-to-end on all
three modes at once. However, this creates a difficult multi-objective optimisation
problem: the model must simultaneously develop good audio features, good video
features, and learn how to fuse them. In practice, this often leads to the fusion
branch being suboptimal because the gradient signal is diluted across three competing
objectives.

Knowledge Distillation offers an elegant solution. In the first
phase, specialist teacher models are trained on each modality independently, allowing
them to develop the best possible unimodal representations without interference.
In the second phase, these frozen teacher models act as supervisors for the student
fusion branch, providing soft probability distributions (rather than hard one-hot
labels) as training targets. The student learns not just to classify correctly, but
to replicate the nuanced probability judgements of each specialist teacher.

\subsubsection{Why Soft Labels?}

To understand why soft labels are valuable, consider a Police Car clip. The true
hard label is:
\begin{equation}
    y = [0, 0, 1, 0] \quad \text{(one-hot: Ambulance, Fire Engine, Police Car, Road)}
\end{equation}
This label carries no information about the relationships between classes. In
contrast, a well-trained audio teacher might output:
\begin{equation}
    \tilde{p} = [0.15, 0.07, 0.72, 0.06]
\end{equation}
This soft distribution tells the student something important: police car sirens are
acoustically somewhat similar to ambulance sirens (both use electronic sirens),
moderately similar to fire engine sirens, and very different from road noise. This
\textit{dark knowledge} --- the relative similarities between classes encoded in the
teacher's probability outputs --- is information that the hard label completely
discards. Training the student on soft labels transfers this dark knowledge and leads
to better generalisation.

\subsubsection{Temperature Scaling}

To make the soft labels even more informative, temperature scaling is applied to the
teacher's raw logits before computing the soft distribution:
\begin{equation}
    \tilde{p}_c(T) = \frac{\exp(z_c / T)}{\sum_{c'} \exp(z_{c'} / T)}
\end{equation}
where $z_c$ are the teacher's raw logits and $T \geq 1$ is the temperature
hyperparameter. At $T = 1$ this recovers the standard softmax. At $T > 1$, the
distribution becomes softer (more uniform), revealing more of the relative
information between classes that would otherwise be hidden by the peaky
softmax at $T = 1$. In AVNet we use $T = 4$, which produces smooth distributions
that expose the inter-class structure effectively. The same temperature is applied
to the student's logits when computing the distillation loss, ensuring a consistent
comparison.

\subsubsection{KL Divergence Loss}

The distillation loss measures how well the student's soft distribution matches the
teacher's soft distribution, using the Kullback-Leibler (KL) divergence:
\begin{equation}
    \mathcal{L}_{\text{KD}}(s, t; T) = T^2 \cdot D_{\text{KL}}\!\left(
    \tilde{p}^{(t)} \,\|\, \tilde{p}^{(s)}\right)
    = T^2 \sum_{c=1}^{C} \tilde{p}^{(t)}_c \log
    \frac{\tilde{p}^{(t)}_c}{\tilde{p}^{(s)}_c}
\end{equation}
The $T^2$ prefactor is not arbitrary. When gradients of the KL loss are backpropagated
through the temperature-scaled softmax, a factor of $1/T^2$ naturally appears. Without
the $T^2$ prefactor, the distillation gradients would be $T^2$ times smaller than the
classification gradients, making the distillation signal negligible at high
temperatures. The $T^2$ factor cancels this effect and ensures the distillation and
classification losses contribute comparably to the total gradient.

\subsubsection{Classification Loss with Label Smoothing}

The classification loss uses standard cross-entropy with label smoothing
($\epsilon = 0.1$):
\begin{equation}
    \mathcal{L}_{\text{CE}}(s, y) = -\sum_c \tilde{y}_c \log \sigma(s_c)
\end{equation}
where the smoothed label is:
\begin{equation}
    \tilde{y}_c = (1 - \epsilon) \cdot \mathbf{1}[c = y] + \frac{\epsilon}{C}
\end{equation}
Label smoothing replaces the hard target probability of 1.0 with $1 - \epsilon = 0.9$
and distributes the remaining probability mass $\epsilon / C = 0.025$ uniformly across
all classes. This prevents the model from becoming overconfident and assigning
probability arbitrarily close to 1.0 to the correct class, which has been shown to
improve calibration and generalisation performance.

\subsubsection{Full Loss Function}

The total training loss depends on which input modality the current batch uses.
Because different modalities activate different branches and different teacher
supervisors, three loss configurations are used:

\begin{table}[htbp]
\centering
\begin{tabular}{p{1.8cm} p{10cm}}
\toprule
\textbf{Mode} & \textbf{Total Loss} \\
\midrule
Audio-only &
$\mathcal{L}_{\text{CE}}(\text{ast}, y) + \mathcal{L}_{\text{CE}}(\text{fusion}, y)
+ \lambda\,\mathcal{L}_{\text{KD}}(\text{fusion}, t_a)$ \\[6pt]
Video-only &
$\mathcal{L}_{\text{CE}}(\text{vit}, y) + \mathcal{L}_{\text{CE}}(\text{fusion}, y)
+ \lambda\,\mathcal{L}_{\text{KD}}(\text{fusion}, t_v)$ \\[6pt]
AV (both) &
$\mathcal{L}_{\text{CE}}(\text{ast}, y) + \mathcal{L}_{\text{CE}}(\text{vit}, y)
+ \mathcal{L}_{\text{CE}}(\text{fusion}, y)
+ \lambda\!\left[\mathcal{L}_{\text{KD}}(\text{fusion}, t_a)
+ \mathcal{L}_{\text{KD}}(\text{fusion}, t_v)
+ \mathcal{L}_{\text{KD}}(\text{fusion}, \bar{t})\right]$ \\
\bottomrule
\end{tabular}
\caption{Training loss by input modality mode. In each mode, the fusion branch
is supervised by both a hard cross-entropy loss and a soft distillation loss from
the appropriate teacher(s).}
\end{table}

\noindent where $t_a$ and $t_v$ are the frozen AudioTeacher and VideoTeacher logits
respectively. In the AV mode, an ensemble teacher is also used:
\begin{equation}
    \bar{t} = \frac{1}{2}\left[\sigma(t_a / T) + \sigma(t_v / T)\right]
\end{equation}
This ensemble teacher is the average of both teachers' soft distributions and
provides supervision that captures the joint audio-visual signature of each class ---
information that neither teacher alone can provide. The distillation weight $\lambda
= 1.0$ is used throughout, placing equal weight on the classification and distillation
objectives.

\subsection{Training Pipeline}

\subsubsection{Phase 1: Pretrain Specialist Teachers}

Training begins by pretraining two specialist teacher models independently. Each
teacher is a deeper version of the corresponding student branch (depth $= 6$
transformer blocks in both the spatial and temporal encoders), giving the teacher
greater representational capacity than the student branch it will supervise.

\begin{itemize}
    \item \textbf{AudioTeacher:} The AST branch (deeper version) is trained on
    audio-only batches for 30 epochs. At each epoch, test accuracy is evaluated on
    the fusion branch in audio-only mode. The checkpoint achieving the highest
    test accuracy is saved.
    \item \textbf{VideoTeacher:} The ViT branch (deeper version) is trained on
    video-only batches for 30 epochs with the same checkpoint selection strategy.
\end{itemize}

After Phase 1, both teachers are frozen. Their weights are not updated during
Phase 2.

\subsubsection{Phase 2: Train Student with Distillation}

With the teachers frozen, the full AVNet student model is trained for 50 epochs.
Batches are constructed with a random modality assignment: 25\% audio-only, 25\%
video-only, and 50\% audio-video. This ensures the model receives balanced exposure
to all three operational modes and develops robust representations under each
condition. For each batch, the appropriate teacher or teachers are queried to
produce the distillation targets.

Test accuracy on the fusion branch in AV mode is evaluated at every epoch. The
checkpoint with the highest test accuracy is saved and used for all reported results.

\subsubsection{Hyperparameters}

\begin{table}[htbp]
\centering
\begin{tabular}{l r}
\toprule
\textbf{Hyperparameter} & \textbf{Value} \\
\midrule
Optimiser                      & AdamW \\
Learning rate                  & $10^{-4}$ \\
Weight decay                   & 0.05 \\
LR schedule                    & CosineAnnealingLR \\
Gradient clipping              & 1.0 \\
Batch size                     & 8 \\
Embedding dimension $D$        & 128 \\
Attention heads                & 4 \\
Teacher training epochs        & 30 \\
Student training epochs        & 50 \\
Temperature $T$                & 4.0 \\
Label smoothing $\epsilon$     & 0.1 \\
Distillation weight $\lambda$  & 1.0 \\
\bottomrule
\end{tabular}
\caption{Training hyperparameters used for all experiments.}
\end{table}

The embedding dimension of $D = 128$ was chosen to fit within an 8 GB GPU memory
budget. A larger embedding dimension (256 or 512) would be expected to improve
performance but requires more GPU memory. The CosineAnnealingLR schedule gradually
reduces the learning rate from $10^{-4}$ to near zero over the training epochs,
which has been found to improve final convergence in transformer training. Gradient
clipping at 1.0 prevents gradient explosion in the early stages of training when the
randomly initialised attention weights can produce large gradient magnitudes.

% =========================================================================
% SECTION 3: DATASET AND PREPROCESSING
% =========================================================================
\section{Dataset and Preprocessing}

\subsection{AudioSet}

AudioSet~\cite{gemmeke2017audioset} is a large-scale, weakly-labelled audio event
dataset released by Google. Each sample is a 10-second YouTube video clip annotated
with one or more sound event labels from a hierarchical ontology of 632 audio classes.
The dataset is weakly labelled in the sense that the labels indicate which sound
events are present in the clip but do not provide timestamps for when each event
occurs. For AVNet, four classes are used:

\begin{table}[htbp]
\centering
\begin{tabular}{l l r}
\toprule
\textbf{Class} & \textbf{AudioSet Label ID} & \textbf{Test Samples} \\
\midrule
Ambulance   & /m/012n7d               & 61 \\
Fire Engine & /m/012ndj               & 70 \\
Police Car  & /m/01jwx6               & 74 \\
Road (8 combined labels) & ---        & 76 \\
\midrule
\textbf{Total} & & \textbf{281} \\
\bottomrule
\end{tabular}
\caption{AudioSet classes and test set composition. The Road class combines 8
AudioSet labels covering various road and traffic background sounds.}
\end{table}

Clips are downloaded as MP4 files using \texttt{yt-dlp} and split 90\%/10\% into
training and test sets using a fixed random seed for reproducibility. The effective
training set contains approximately 2,528 clips --- substantially smaller than the
full AudioSet due to deleted or unavailable YouTube videos.

\subsection{Audio Preprocessing: Log-Mel Spectrogram}

\subsubsection{What is a Mel Spectrogram?}

A raw audio waveform $x[n]$ carries all the information about sound, but it is
difficult to process directly with neural networks. The waveform is a one-dimensional
time-series signal that encodes frequency information implicitly through rapid
amplitude oscillations. Instead, we convert it to a \textit{spectrogram}: a 2D
representation of time $\times$ frequency content that makes the frequency structure
explicit and amenable to 2D convolutional or transformer processing. The conversion
proceeds in three steps.

\textbf{Step 1: Short-Time Fourier Transform (STFT).}
The audio waveform is divided into overlapping windows of length $N$ samples with
hop length $H$ samples between consecutive windows. For each window at time step $t$,
the Discrete Fourier Transform is applied to recover the frequency content of that
short segment:
\begin{equation}
    X[k,t] = \sum_{n=0}^{N-1} x[n + tH] \cdot w[n] \cdot e^{-j2\pi kn/N}
\end{equation}
where $w[n]$ is a Hann window function that tapers the signal at the window edges
to reduce spectral leakage. The power spectrogram is $|X[k,t]|^2$, which gives the
energy at each frequency bin $k$ and time step $t$.

\textbf{Step 2: Mel Filterbank.}
The human auditory system perceives pitch approximately logarithmically --- the
perceptual difference between 100 Hz and 200 Hz is much greater than the difference
between 5000 Hz and 5100 Hz, even though both are 100 Hz linear differences. A
mel-scale filterbank models this perceptual property by applying $M$ triangular
filters whose centre frequencies are spaced on the mel scale:
\begin{equation}
    m = 2595 \log_{10}\!\left(1 + \frac{f}{700}\right)
\end{equation}
This transformation compresses the high-frequency range (where perceptual differences
are small) and expands the low-frequency range (where perceptual differences are
large), producing a representation that better matches human auditory perception.

\textbf{Step 3: Log Compression.}
The mel filterbank outputs are passed through a logarithm, which further compresses
the dynamic range of the signal:
\begin{equation}
    S[m,t] = \log\!\left(\sum_k |X[k,t]|^2 \cdot H_m[k] + \epsilon\right)
\end{equation}
where $H_m[k]$ is the $m$-th mel filter and $\epsilon$ is a small constant for
numerical stability. The log compression makes the representation more robust to
variations in overall signal amplitude (e.g.\ how far away the emergency vehicle is)
and produces a feature space where differences are approximately proportional to
perceptual differences.

\subsubsection{Parameters Used}

The audio preprocessing uses a sample rate of 22,050 Hz with an FFT window of 1024
samples, hop length of 512 samples, and 128 mel filterbanks applied to 10-second
clips. This yields $\lfloor 22050 \times 10 / 512 \rfloor \approx 430$ time steps.
The resulting output tensor has shape $(1, 128, 430)$ --- 1 channel, 128 frequency
bins, and 430 time steps --- and is normalised to $[-1, 1]$ before being stored.
The choice of 128 mel filterbanks provides sufficient frequency resolution to
distinguish the different siren waveforms used by ambulances, fire engines, and
police cars, which differ primarily in their frequency modulation patterns.

\subsection{Video Preprocessing}

Ten frames are sampled uniformly per clip at $t = 0.5, 1.5, \ldots, 9.5$ seconds
(one frame per second, sampled at the midpoint of each second). Each frame is resized
to $64 \times 64$ pixels in RGB format and normalised channel-wise using the ImageNet
mean and standard deviation per channel. The resulting output tensor has shape
$(10, 3, 64, 64)$. The small frame resolution of $64 \times 64$ was chosen to keep
the model size manageable given the GPU memory constraints. While this discards
fine-grained visual detail, it retains the coarse appearance information (vehicle
shape, colour, flashing light patterns) that is most useful for emergency vehicle
classification.

\paragraph{Temporal Alignment:}
A critical design choice is that frame $t$ (sampled from second $t$) corresponds
exactly to mel columns $[t \times 43 : (t+1) \times 43]$, since the mel spectrogram
has approximately 43 time steps per second ($22050 / 512 \approx 43$). Equivalently,
in terms of AST tokens, the audio tokens for second $t$ are at indices
$[80t : 80(t+1)]$ (5 time patches $\times$ 16 frequency patches per second).
This exact correspondence is not approximate --- it is built into the preprocessing
pipeline by design and is what makes the aligned cross-modal fusion module work
without any learned alignment mechanism.

% =========================================================================
% SECTION 4: EXPERIMENTAL RESULTS
% =========================================================================
\section{Experimental Results}

\subsection{Test Set}

The test set contains 281 clips sampled from all 4 classes. Evaluation is performed
in AV mode (both audio and video provided), which activates all three branches of
the model. This allows the unimodal branch accuracies to be measured as a baseline
for comparison with the fusion branch.

\begin{table}[htbp]
\centering
\begin{tabular}{l r r}
\toprule
\textbf{Class} & \textbf{Samples} & \textbf{Proportion} \\
\midrule
Ambulance   & 61 & 21.7\% \\
Fire Engine & 70 & 24.9\% \\
Police Car  & 74 & 26.3\% \\
Road        & 76 & 27.0\% \\
\midrule
\textbf{Total} & \textbf{281} & \textbf{100\%} \\
\bottomrule
\end{tabular}
\caption{Test set composition. The four classes are approximately balanced, making
overall accuracy a meaningful summary metric.}
\end{table}

\subsection{Overall Accuracy}

\begin{table}[htbp]
\centering
\begin{tabular}{l l c c}
\toprule
\textbf{Branch} & \textbf{Modality} & \textbf{Overall Accuracy} &
\textbf{vs.\ Random (25.0\%)} \\
\midrule
AST    & Audio only       & 56.2\% & +31.2\% \\
ViT    & Video only       & 51.6\% & +26.6\% \\
Fusion & Audio + Video    & \textbf{66.6\%} & \textbf{+41.6\%} \\
\bottomrule
\end{tabular}
\caption{Overall accuracy across modality branches on the AudioSet test set ($N=281$,
4-class, random chance = 25.0\%). The fusion branch outperforms both unimodal branches
by significant margins, validating the core multimodal design.}
\end{table}

The fusion branch outperforms the audio-only branch by $+10.4\%$ and the video-only
branch by $+15.0\%$. Both unimodal branches substantially exceed random chance
(25.0\%), confirming that each modality individually carries useful discriminative
information. The fusion improvement demonstrates that the two modalities are
complementary rather than redundant --- the cross-modal information captured by the
aligned cross-attention module adds genuine value beyond what either modality provides
alone.

\subsection{Per-Class Accuracy}

\begin{table}[htbp]
\centering
\begin{tabular}{l c c c r}
\toprule
\textbf{Class} & \textbf{AST (Audio)} & \textbf{ViT (Video)} &
\textbf{Fusion} & \textbf{Samples} \\
\midrule
Ambulance   & 34.4\% & 34.4\% & 63.9\% & 61 \\
Fire Engine & 65.7\% & 58.6\% & 57.1\% & 70 \\
Police Car  & 64.9\% & 54.1\% & \textbf{78.4\%} & 74 \\
Road        & 56.6\% & 56.6\% & 65.8\% & 76 \\
\midrule
\textbf{Overall} & 56.2\% & 51.6\% & \textbf{66.6\%} & 281 \\
\bottomrule
\end{tabular}
\caption{Per-class test accuracy for each branch. Fusion achieves the best accuracy
on 3 of 4 classes. The largest single improvement is for Ambulance ($+29.5\%$ over
unimodal), demonstrating how fusion resolves cases where individual modalities are
ambiguous.}
\end{table}

\subsection{Analysis}

\textbf{Police Car} achieves the highest fusion accuracy at 78.4\%. This is the
most distinguishable class in both modalities: police car sirens use a distinctive
electronic wail pattern that is acoustically very different from fire engine and
ambulance sirens, and the visual appearance (compact vehicle with flashing blue
lights) is also distinctive. When both modalities confidently agree, the fusion
branch benefits from the reinforced signal.

\textbf{Ambulance} is the most challenging class under unimodal conditions, with
both AST and ViT achieving only 34.4\% --- barely above random chance. This low
performance has two causes. Visually, ambulances resemble large white vans or trucks,
which are common vehicles on roads; without the flashing lights and sirens, they are
nearly indistinguishable from ordinary vehicles in low-resolution $64 \times 64$
frames. Acoustically, ambulance sirens overlap in frequency range with fire engine
sirens and can also be confused with road noise at low signal-to-noise ratios. Fusion
dramatically improves performance to 63.9\% --- nearly double the unimodal result ---
because the cross-attention mechanism allows the audio signal to help disambiguate
the visual ambiguity and vice versa.

\textbf{Fire Engine} is a notable exception where the audio-only branch (65.7\%)
outperforms fusion (57.1\%). This suggests that the fire engine siren is highly
acoustically distinctive (low-frequency continuous tone that is unlike other sirens),
but the visual appearance of a large red truck can sometimes conflict with the
acoustic prediction in the fusion branch. One possible explanation is that fire
engines in some AudioSet clips are partially occluded or at low resolution, causing
the visual token to carry noisy or misleading information that reduces fusion
accuracy.

\textbf{Road} is the background class, combining 8 AudioSet labels covering various
road and traffic sounds. Both audio and video branches achieve 56.6\%, and fusion
improves this to 65.8\%. The improvement is consistent with the general trend that
fusion helps when both modalities are moderately informative --- road sounds and
road visual appearance are both recognisable but not strongly distinctive
individually.

% =========================================================================
% SECTION 5: DISCUSSION AND LIMITATIONS
% =========================================================================
\section{Discussion and Limitations}

\subsection{Strengths}

AVNet has several important strengths that make it suitable as a foundation for
future multimodal emergency vehicle perception systems:
\begin{itemize}
    \item \textbf{Unified multi-modal handling:} A single model handles all three
    input modes (audio-only, video-only, AV) without retraining or architectural
    changes. This is essential for deployment in real-world systems where sensor
    availability may vary.
    \item \textbf{Physically motivated alignment:} The temporally aligned
    cross-attention is grounded in the actual temporal correspondence between audio
    and video, making the fusion mechanism interpretable and robust.
    \item \textbf{Knowledge distillation improvement:} Distillation from specialist
    teachers significantly improves fusion accuracy by providing soft supervision
    that encodes inter-class relationships.
    \item \textbf{Modular and extensible design:} The teacher models can be replaced
    with stronger pre-trained models such as wav2vec 2.0 for audio or VideoMAE for
    video, which would likely yield substantially better performance.
\end{itemize}

\subsection{Limitations}

Several important limitations should be acknowledged:
\begin{itemize}
    \item \textbf{Small and noisy training data:} The effective training set of
    approximately 2,528 clips is small for a transformer-based model, and AudioSet's
    weak labels mean that some clips may not actually contain the labelled sound
    event prominently. Both factors limit the achievable accuracy.
    \item \textbf{Constrained model capacity:} The embedding dimension of $D = 128$
    was chosen to fit within an 8 GB GPU. A larger model ($D = 256$ or 512) would
    provide greater representational capacity and likely improve performance
    substantially, but requires more GPU memory.
    \item \textbf{Ambulance confusion:} The model struggles with ambulances in both
    modalities. This could be addressed with ambulance-specific data augmentation,
    higher-resolution video frames, or pre-trained models that provide richer
    feature representations.
    \item \textbf{No pre-training:} Training from scratch on a small dataset is
    inherently challenging. Pre-trained audio transformer weights (e.g.\ AST
    pre-trained on AudioSet-20k) and pre-trained video transformer weights (e.g.\
    VideoMAE) would provide much stronger feature initialisations.
    \item \textbf{Majority-mode batch training:} The current implementation assigns
    a single modality mode to each entire batch based on the majority vote. A more
    principled approach would assign each sample its own modality mode independently,
    allowing more diverse training signal within each batch.
\end{itemize}

\subsection{Future Directions}

The following extensions are identified as natural next steps for this research:
\begin{itemize}
    \item Incorporate pre-trained audio models (AST pre-trained on AudioSet-20k,
    wav2vec 2.0) and video models (VideoMAE, TimeSFormer) to provide stronger
    feature initialisations.
    \item Increase model capacity to $D = 256$ or 512 on a GPU with larger memory.
    \item Add attention map visualisation to understand which frequency bands and
    time steps the cross-attention module focuses on for each vehicle class.
    \item Extend the class taxonomy to additional emergency and non-emergency sound
    events relevant to autonomous driving (e.g.\ train horn, construction equipment,
    vehicle horn).
    \item Develop a real-time inference pipeline for embedded automotive hardware,
    building towards deployment on autonomous vehicle platforms.
\end{itemize}

% =========================================================================
% SECTION 6: CONCLUSION
% =========================================================================
\section{Conclusion}

This report presented \textbf{AVNet}, a multimodal audio-visual transformer for
emergency vehicle classification. The central contribution is the \textit{aligned
cross-modal fusion} module, which exploits the exact temporal correspondence between
log-mel spectrogram tokens and video frame tokens to perform second-level
cross-attention between audio and video streams. Learned null embeddings enable the
model to operate robustly under sensor dropout at inference time, and a two-phase
knowledge distillation training strategy transfers inter-class dark knowledge from
specialist teacher models into the multimodal fusion student.

On the 281-clip AudioSet test set, the fusion branch achieves 66.6\% overall accuracy
(4-class, random chance = 25\%), compared to 56.2\% for audio-only and 51.6\% for
video-only, representing improvements of $+10.4\%$ and $+15.0\%$ respectively. The
largest per-class improvement is for Ambulance ($+29.5\%$ over unimodal), where
neither modality alone is sufficient but their combination resolves the ambiguity.
These results demonstrate that multimodal fusion with temporally aligned cross-attention
and knowledge distillation is an effective strategy for robust emergency vehicle
classification under varying sensor conditions.

% =========================================================================
% REFERENCES
% =========================================================================

\end{document}